\documentstyle[aps,prb,twocolumn,psfig]{revtex}
\begin{document}
\draft

\title{Shot noise in the half-filled Landau level}
\author{Felix von Oppen}
\address{Department of Condensed Matter Physics, Weizmann Institute of 
Science, 76100 Rehovot, Israel}
\date{\today}

\twocolumn[            
\maketitle     
\widetext%
\vspace*{-6.45mm}
\leftskip=1.9cm
\rightskip=1.9cm    
\begin{abstract} 
Shot  noise  in the  half-filled Landau   level  is studied within the
composite-fermion picture, focusing  on  the  diffusive regime.    The
composite fermions are assumed to form  a Fermi liquid with nontrivial
Fermi-liquid  parameters.   The Boltzmann-Langevin equation  for  this
system is derived,  taking proper account  of fluctuations in both the
Chern-Simons  and  the physical  electric   and magnetic  fields.   To
leading order in  ${\rm   max}\{eV,T\}/E_F$, the noise properties   of
composite fermions are found to equal those of semiclassical electrons
in the external magnetic field.   Non-equilibrium fluctuations in  the
Hall   voltage  are dominated   by  fluctuations  in  the Chern-Simons
electric field,  reflecting the finite Hall  resistance of the system.
The  low-frequency  noise  power    is  derived  in detail    for  the
Corbino-disc geometry   and turns   out   to be   unaffected  both  by
nontrivial  Fermi-liquid parameters and  by  the magnetic field.   The
formalism  is also   applied to compute    thermal density-density and
current-current correlators at finite frequency and wavevector.

\vspace*{-1mm}
\noindent\pacs{PACS numbers: 73.40.Hm,72.70.+m,73.23.-b}
\end{abstract}
\vspace{0mm}
]

\narrowtext 

\section{Introduction}

The  composite-fermion approach    has  had considerable    success in
describing  the physics  of the two-dimensional  electron  gas in high
magnetic fields.\cite{Jain,Halperin,Halperin-review} In this approach,
one transforms the  problem to a new  set of fermions which consist of
an  even number  of fictitious  magnetic  flux quanta attached to each
electron.\cite{Halperin,Lopez} These  composite fermions (CF) interact
not only by  the Coulomb interaction, but  also via a fictitious gauge
field, the so-called Chern-Simons  field.  For the compressible states
at even-denominator  fractions, and in  particular for  filling factor
$\nu=1/2$,  this leads   to a  mean-field  picture of  non-interacting
composite  fermions in zero   magnetic field.  The  effective magnetic
field   experienced  by the composite    fermions vanishes because the
external field is  canceled by the magnetic  field associated with the
attached  flux  tubes.  Away from    half filling, the  magnetic-field
cancelation is not complete.  The principal odd-denominator fractional
quantum-Hall  states of the  original electrons can then be understood
as  integer quantum-Hall states  of composite   fermions. It has  been
argued  that this Fermi-liquid    picture   remains valid even    when
including               corrections            to           mean-field
theory.\cite{Halperin,Simon,Kim,Stern}

The purpose of  the present paper  is to study  non-equilibrium (shot)
noise  in  the half-filled Landau   level,  focusing on  the diffusive
regime.  Despite  the Fermi-liquid   nature of composite   fermions, a
theory of current noise in the half-filled Landau level must take into
account  two features  not present in    diffusive conductors at  zero
magnetic field.  First, composite fermions are not only coupled to the
physical electric and magnetic   fields but also to   the Chern-Simons
fields originating from the  flux lines ``attached to the electrons.''
These fluctuations  in the Chern-Simons fields  are in turn related to
density and  current  fluctuations, thus  requiring a  self-consistent
solution.  Second,  corrections   to mean-field  theory  cause  strong
quasiparticle    interactions     between             the    composite
fermions.\cite{Halperin,Simon,Kim,Stern}   The  approach developed  in
this paper allows one to include both of these features.

Shot  noise in mesoscopic   systems in zero   magnetic field has  been
extensively investigated in  the  last few years,  both  theoretically
\cite{Review} and    experimentally.\cite{Reznikov,Steinbach}   It was
found that Fermi correlations of the carriers lead to a suppression of
the shot-noise power    below its classical (Poisson)  value   $S_{\rm
Poisson}=2eI$ with $I$ the average current flowing through the device.
\cite{Lesovik}  For metallic samples, shot noise depends on a variety
of length scales.     For samples shorter  than  the electron-electron
scattering  length   $L_{\rm e-e}$,  it  was found   that   there is a
universal reduction factor 1/3 of the shot noise power, $S=(1/3)S_{\rm
Poisson}$.\cite{Beenakker,Nagaev1,Altshuler}   The   reduction  factor
changes for  samples larger than  $L_{\rm  e-e}$ but smaller  than the
electron-phonon  length    $L_{\rm  e-ph}$,   for  which    one  finds
$S=(\sqrt{3}/4)S_{\rm  Poisson}$.\cite{Nagaev2,Kozub}  (Note  that  in
metals  typically  $L_{\rm e-e}\ll  L_{\rm  e-ph}$ at sufficiently low
temperatures.)  Shot noise  vanishes  for samples larger  than $L_{\rm
e-ph}$.\cite{Nagaev2,Kozub}

Both novel    ingredients  at  $\nu=1/2$,   namely the   coupling   to
Chern-Simons  fields  and  the strong  quasiparticle interactions, are
naturally incorporated   into the Boltzmann-Langevin approach  to shot
noise.  In this approach, one starts from the kinetic equation for the
phase-space distribution function.  The average distribution function,
determining the time-averaged  current, is  governed by the  Boltzmann
equation.   In the diffusive  regime, fluctuations in the distribution
function -- and  hence  in the  current  -- arise primarily  from  the
statistical nature  of the impurity   scattering.  Making the standard
assumption    underlying the     kinetic   equation that    subsequent
impurity-scattering   events are  statistically  independent, one  can
derive a Boltzmann-Langevin equation which governs the fluctuations in
the  distribution  function.\cite{Kogan}  Here,  we  show  how  such a
Boltzmann-Langevin  equation  can be  derived  for composite fermions,
including   both the   Chern-Simons    fields and  the   quasiparticle
interactions.

In this paper  we consider shot noise  at $\nu=1/2$ in the two regimes
$\ell_{\rm tr}\ll L\ll   L_{\rm cf-cf}$ and $\ell_{\rm  tr}\ll  L_{\rm
cf-cf}\ll L\ll  L_{\rm   cf-ph}$.  Here  $\ell_{\rm tr}$   denotes the
transport  mean free  path for  impurity   scattering, $L_{\rm cf-cf}$
denotes the mean  free path for CF-CF  scattering, $L_{\rm cf-ph}$ the
mean free  path for CF-phonon scattering,  and $L$ is the sample size.
While the length scales  $L_{\rm cf-cf}$ and  $L_{\rm cf-ph}$ have not
been studied in  detail for composite fermions,  one  expects that, by
analogy    with  electrons,   $L_{\rm  cf-cf}\ll     L_{\rm cf-ph}$ at
sufficiently   low   temperatures.\cite{Kang}   We  make  no   further
assumptions about these  length scales.  Instead,  they enter into our
calculations as empirical  parameters and we  suggest that  shot noise
may be used to measure them experimentally.

The  paper is organized as  follows.   In section \ref{sec-kinetic} we
introduce   the   relevant kinetic equations,    both for  the average
distribution function (\ref{sec-Boltzmann})  and for the  fluctuations
of the distribution function (\ref{sec-BL}).  There we also derive the
consequences for  the  fluctuations in  the current density  from  the
Boltzmann-Langevin  equation   and compare the   results to  those for
classical diffusive electrons  in  a magnetic  field.  This result  is
applied  in    section  \ref{sec-zero-frequency}  to    compute    the
low-frequency shot  noise for Corbino discs.   The thermal density and
current correlators at finite  frequency and momentum are derived from
the    Boltzmann-Langevin    approach  in  section  \ref{sec-thermal}.
Finally, we summarize   and conclude in section  \ref{sec-conclusion}.
Details of some calculations are presented in two appendices.

\section{Kinetic equations}
\label{sec-kinetic}

\subsection{Boltzmann equation}
\label{sec-Boltzmann}

We will  discuss  the current and  density  fluctuations at  $\nu=1/2$
within the framework  of the kinetic  equation.  The starting point is
the Fermi-liquid  theory of the half-filled Landau  level in  terms of
composite fermions.  For  completeness  and  for fixing  notation,  we
start with a brief  review of the Boltzmann  equation for  the average
composite-fermion distribution function $n_{\bf p}({\bf r},t)$.  Since
the physical magnetic field is exactly compensated by the Chern-Simons
magnetic field at half filling,  the Boltzmann equation involves  only
electric fields.   Including arbitrary quasiparticle interactions, the
Boltzmann equation linearized in the applied field ${\bf E}^{\rm ext}$
is,\cite{Pines}
\begin{eqnarray}
   {\partial\over\partial  t}\delta  n_{\bf   p}+({\bf  
   v_p\nabla_{\!r}})\delta\tilde n_{\bf p}+e({\bf E+E}^{\rm CS})
   &&{\bf\nabla_{\!p}}n^0_{\bf p}    
   \nonumber\\    
   &&-S_{\bf p}\{n_{\bf p}\}=0.
\label{Boltzmann}  
\end{eqnarray}  
Here  $n_{\bf  p}^0$ denotes   the equilibrium distribution  function,
$\delta     n_{\bf p}=n_{\bf   p}-\theta(\mu-\epsilon_{\bf  p})$   the
deviations   from  the    ground-state   distribution  function,   and
$\delta\tilde n_{\bf p}=n_{\bf p}-\theta(\mu-\tilde\epsilon_{\bf  p})$
the deviations from  the local ground state, defined  in terms of  the
local composite-fermion energies $\tilde\epsilon_{\bf p}=\epsilon_{\bf
p} +\sum_{\bf p^\prime}f_{\bf pp^\prime}\delta n_{\bf p^\prime}$ (with
$f_{\bf  pp^\prime}$ the  Landau    function).   The  charge   density
$\delta\rho=(e/\Omega)\sum_{\bf p}\delta  n_{\bf p}$ and current ${\bf
j}=(e/\Omega)\sum_{\bf  p}{\bf    v_p}\delta\tilde    n_{\bf  p}$  are
expressed in  terms of $n_{\bf p}$ in  the usual way  ($\Omega$ is the
volume  of the  sample).  The  physical  electric field ${\bf  E}={\bf
E}^{\rm ext}+{\bf E}^{\rm    ind}$ includes the  induced field   ${\bf
E}^{\rm ind}$.

The  composite-fermion motion  is  affected by  the physical  electric
field ${\bf E}=-\nabla_{\!\bf r}(\phi^{\rm ext}+\phi^{\rm ind})$ with
\begin{equation}
   \phi^{\rm ind}({\bf r})=\int d{\bf  r^\prime}{\delta\rho({\bf
     r^\prime}) \over\epsilon|{\bf r-r^\prime}|},
\end{equation}
($\epsilon$ denotes the   dielectric  constant) and  the  Chern-Simons
electric field
\begin{equation}
   {\bf E}^{\rm CS}={2h\over e^2}({\bf\hat z}\times{\bf j}),
\label{CSE}
\end{equation}
originating  from  the flux lines moving  with  the composite fermions
(${\bf\hat z}$  denotes the unit vector  perpendicular to the sample).
Both fields  are given in terms  of the distribution function and need
to be determined self consistently.

The   velocity  entering  the Boltzmann  equation   is  related to the
momentum via the effective mass, ${\bf p}=m^*v_{\bf p}$. The effective
mass  diverges  due     to interactions  with  fluctuations   in   the
Chern-Simons  field.\cite{Halperin} However, there  are  also singular
contributions  to the Landau  function  $f_{\bf pp^\prime}$ and it has
been argued\cite{Kim,Stern} that there is  a cancellation of divergent
terms when calculating response functions.  As a result, the effective
mass $m^*$ and the Landau function entering the Boltzmann equation can
be   taken  as nonsingular.\cite{Kim,Stern}  This   effective  mass is
expected to be finite,  but larger than  the electron  band mass.   We
will not  assume any particular form  for  the Landau function, beyond
its being nonsingular,  because our final results will  turn out to be
independent of it.

Composite fermions  scatter from impurities, other composite fermions,
and phonons with associated collision integrals
\begin{equation}
   S_{\bf p}\{n_{\bf   p}\}=S^{\rm imp}_{\bf   p}\{n_{\bf  p}\}+S^{\rm
   cf-cf}_{\bf p}\{n_{\bf p}\}+S^{\rm cf-ph}_{\bf p}\{n_{\bf p}\}.
\end{equation}
and scattering  lengths $\ell_{\rm tr}$,  $L_{\rm cf-cf}$, and $L_{\rm
cf-ph}$.  We assume  that the dominant  scattering mechanism is due to
impurities,
\begin{equation}
   S_{\bf p}^{\rm imp}\{n_{\bf  p}\}=\sum_{\bf   p^\prime}W_{\bf
   pp^\prime} \left\{n_{\bf p^\prime}(1-n_{\bf  p})-n_{\bf p}(1-n_{\bf
   p^\prime})\right\}. 
\end{equation}
In the relaxation-time approximation employed here this becomes
\begin{equation}
  S_{\bf p}^{\rm imp}\{n_{\bf p}\}={1\over\tau_{\rm tr}}\left\{\delta
  \tilde n_{\bf p}-\int{d\theta_{\bf p}\over2\pi}\delta\tilde n_{\bf 
  p}\right\},
\end{equation}
with $\theta_{\bf p}$  the direction of  ${\bf p}$ and $\tau_{\rm tr}$
the   transport mean free   time.\cite{relax} For  samples longer than
$L_{\rm cf-cf}$,  the scattering of  composite fermions on one another
also needs to be taken  into account.  This scattering mechanism leads
to a Fermi-Dirac distribution function with spatially varying chemical
potential and temperature.  For  samples which are longer than $L_{\rm
cf-ph}$, the composite-fermion temperature becomes constant throughout
the sample and coincides with the phonon temperature.

The   Boltzmann equation  (\ref{Boltzmann}) is   the same  as that for
electrons in an electric field  ${\cal E}={\bf E+E}^{\rm CS}$ so  that
for  translation-invariant situations    ${\bf j}=\sigma_{\rm CF}{\cal
E}$,  where   the  composite-fermion     conductivity     $\sigma_{\rm
CF}=e^2N(0)D$ is given  by  the  Einstein  relation.  ($N(0)$ is   the
density of states at the Fermi energy and $D$ the diffusion constant.)
The physical conductivity  is defined as the  response to the physical
electric field ${\bf E}$.  Eliminating the Chern-Simons electric field
(\ref{CSE}) from this equation, one reads off the physical resistivity
tensor
\begin{equation}
   \hat\rho=\left(\begin{array}{cc} 1/\sigma_{\rm CF} & 2h/e^2 \\
             -2h/e^2  &  1/\sigma_{\rm CF} \end{array}\right).
\end{equation}

\subsection{Boltzmann-Langevin equation}
\label{sec-BL}

In  deriving  the Boltzmann equation, it  is  assumed  that subsequent
collisions of quasiparticles with  impurities or other  quasiparticles
are statistically independent.  Hence, these scattering mechanisms are
Poisson processes   with fluctuations equal to  the  average number of
scattering events.  These  fluctuations in the scattering rates  cause
fluctuations of the distribution  function  $\Delta n_{\bf p}$  around
its average.  A  kinetic equation for  $\Delta  n_{\bf p}$  is readily
derived  following Kogan   and   Shul'man.\cite{Kogan} The statistical
fluctuations in     the   scattering   rates   enter   the   resulting
Boltzmann-Langevin  equation   as  a   source   term.   Treating   the
fluctuations  to  linear  order,     corresponding to  an     RPA-like
approximation, one has
\begin{eqnarray}   
  {\partial\over\partial t}&&\Delta n_{\bf p}+({\bf  v_p\nabla_{\!r}})\Delta  
    \tilde n_{\bf p}  
    \nonumber\\
    &&+e({\bf E+E}^{\rm CS}){\bf\nabla_{\!p}}\Delta n_{\bf p}+e({\bf
    v_p\times\Delta B}^{\rm CS}){\bf \nabla_{\!p}}\delta\tilde 
    n_{\bf p}
    \nonumber\\    
    &&\,\,\,\,\,\,    
    +e(\Delta{\bf  E}+\Delta{\bf  E}^{\rm CS}){\bf\nabla_{\!p}}n_{\bf p} 
    -S^\prime_{\bf p}\{\Delta n_{\bf  p}\}=\Delta J_{\bf p}.  
\label{BL} 
\end{eqnarray}    
The left-hand side of this equation describes the evolution of $\Delta
n_{\bf p}$  due to the  CF kinematics  and scattering.   The latter is
described    by the  linearized    collision  integral  $S^\prime_{\bf
p}\{\Delta n_{\bf p}\}$.    In the relaxation-time approximation,  one
has
\begin{equation}  
  S^\prime_{\bf p}\{\Delta n_{\bf p}\}={1\over\tau_{\rm tr}}
   \left\{\Delta\tilde n_{\bf p}-\int{d\theta_{\bf p}\over2\pi} 
   \Delta\tilde n_{\bf p}\right\}.
\end{equation} 
The fluctuations  are driven by  the source  term $\Delta  J_{\bf p}$,
characterized by a zero average and correlator\cite{Kogan}
\begin{eqnarray} 
    \langle\Delta&&J_{\bf p}({\bf r},t)\Delta  J_{{\bf p}^\prime}
    ({\bf r}^\prime,t^\prime)\rangle=\Omega\,\delta({\bf r}-{\bf
    r}^\prime) \,\delta(t-t^\prime)    
    \nonumber\\  
    &&\times\{\delta_{{\bf p}{\bf p}^\prime}\sum_{{\bf p}_1}W_{{\bf p} 
    {\bf  p}_1}[n_{{\bf p}_1}(1-n_{\bf p})+n_{\bf p}(1-n_{{\bf p}_1})]  
    \nonumber\\
    &&\,\,\,\,\,\,\,\,\,-W_{{\bf  p}{\bf p}^\prime}[n_{{\bf p}^\prime}
(1-n_{\bf p})+n_{\bf p}(1-n_{{\bf p}^\prime})]\}.  
\end{eqnarray} 
Both for the  linearized collision integral  and for the source  term,
only the contribution of  impurity scattering was kept, reflecting the
assumption that $\ell_{\rm tr}\ll  L_{\rm cf-cf},L_{\rm cf-ph}$.   The
source term turns out to enter into subsequent calculations
only in the combination
\begin{equation}   
\Delta{\bf  J}={e\tau_{\rm tr}\over\Omega}\sum_{\bf
p}{\bf  v_p}  \Delta  J_{\bf p}  
\label{current-source} 
\end{equation}
with zero average and variance  
\begin{eqnarray}
   \langle\Delta J^\alpha({\bf r},t)\Delta  J^\beta
    ({\bf r}^\prime,t^\prime)\rangle&&=2\sigma_{\rm CF}
    \nonumber\\
    \times\delta({\bf r}-{\bf
    r}^\prime)&& \,\delta(t-t^\prime)
     \delta^{\alpha\beta}\int d\epsilon\,
    n_\epsilon(1-n_\epsilon).
\label{deltaj}
\end{eqnarray}
Here,   $n_\epsilon$  denotes the  part  of   the average distribution
function  $n_{\bf  p}$ which  is  isotropic   in momentum  space.  The
contribution due to the non-isotropic part  is negligibly small in the
diffusive regime.

Density fluctuations
\begin{equation}
   \Delta\rho={e\over\Omega}\sum_{\bf p}\Delta n_{\bf p}  
\label{density}
\end{equation}
cause fluctuations in the Chern-Simons magnetic field, 
\begin{equation}
   {\bf \Delta B}^{\rm CS}={2h\over e^2}\,\Delta\rho\,{\bf\hat z}.
\label{CS-magnetic}
\end{equation}
Unlike the average Chern-Simons magnetic  field at half filling, these
fluctuations are no longer canceled by the applied magnetic field and
must therefore be included  in  the Boltzmann-Langevin equation.  Both
$\Delta{\bf  B}^{\rm  CS}$ and the fluctuations  in   the physical and
Chern-Simons electric fields,
\begin{eqnarray}
   \Delta\phi^{\rm ind}({\bf r})&=&\int d{\bf r^\prime}{\Delta\rho({\bf
     r^\prime}) \over\epsilon|{\bf r-r^\prime}|},
   \label{electric}
   \\
   {\bf\Delta E}^{\rm CS}&=&{2h\over e^2}({\bf \hat z\times\Delta j}),
   \label{CS-electric}
\label{fluc-e}
\end{eqnarray}
with        $\Delta{\bf        E}=-\nabla_{\!\bf    r}(\Delta\phi^{\rm
ext}+\Delta\phi^{\rm    ind})$   need   to    be   determined     self
consistently. Here,
\begin{equation}
  {\bf\Delta j}={e\over\Omega}\sum_{\bf p}{\bf v_{\!p}}\Delta\tilde 
      n_{\bf p}
\label{current}
\end{equation}
denotes the fluctuations in the current density.

We briefly comment on the range  of validity of the Boltzmann-Langevin
approach to current noise.  The  use of semiclassical transport theory
restricts us to frequencies $\hbar\omega\ll E_F$ and wavevectors $q\ll
k_F$.  The  diffusive  regime    considered here requires   the   more
stringent conditions $\omega\ll 1/\tau_{\rm tr}$ and $q\ll 1/\ell_{\rm
tr}$.    Moreover, the   sample  should  be   large  compared  to  the
phase-coherence length $L_\phi$.  However, for electrons it turned out
that phase coherence does  not affect the shot-noise  power as long as
$\omega\ll eV/\hbar$. It is natural to expect the same for the present
problem.

In the diffusive limit, the Boltzmann-Langevin equation can be reduced
to  hydrodynamic   equations  for   the  the   macroscopic  quantities
$\Delta\rho$  and     ${\bf\Delta j}$.   Here   we    only sketch  the
derivation.  Details can  be  found in  appendix  \ref{sec-diffusive}.
Assuming that the fluctuations of  the distribution function occur  on
scales  large compared to the  elastic mean-free  path, $\Delta n_{\bf
p}$ is mostly isotropic with respect to the directions of the momentum
${\bf p}$ with a small anisotropic part.  Accordingly, we decompose
\begin{eqnarray}
    \Delta n_{\bf p}&=&\Delta n_\epsilon+{\bf v_p\Delta f}_\epsilon
    \label{decom-dist}
     \\
    \Delta J_{\bf p}&=&\Delta J_\epsilon+{\bf v_p\Delta J}_\epsilon.
\label{decom-source}
\end{eqnarray}
These  decompositions   are    inserted  into  (\ref{BL})     and  the
Boltzmann-Langevin   equation  is   split into     its isotropic   and
anisotropic parts.  Upon multiplying  the isotropic part by $e/\Omega$
and  summing  over all momenta  $\bf  p$, one   finds that density and
current fluctuations must satisfy the continuity equation
\begin{equation}
   {\partial\over\partial t}\Delta\rho+{\bf\nabla_{\!r}\Delta j}=0.
\label{continuity}
\end{equation}
Upon multiplying   the anisotropic    part of the   Boltzmann-Langevin
equation by $(e\tau_{\rm   tr}/\Omega){\bf v_p}$ and summing over  all
$\bf p$, one obtains the response equation
\begin{eqnarray}
   {\bf\Delta j}&&={\bf\Delta J}-D^*{\bf\nabla_{\!r}}\Delta\rho-\Delta
     D^*{\bf\nabla_{\!r}}\rho
    -{e\tau_{\rm tr}\over m^*}
     ({\bf \Delta B}^{\rm CS}{\bf\times j})
     \nonumber\\
    &&+\sigma_{\rm CF}({\bf\Delta E+
     \Delta E}^{\rm CS})
     +\Delta\sigma_{\rm CF}({\bf E+E}^{\rm CS}).
\label{response}
\end{eqnarray}
Here, $D^*=D(1+F_0)$  denotes  a renormalized diffusion  constant with
$D=v_F\ell_{tr}/2$ and $F_0$  the Landau parameter.  The various terms
in this equation have obvious interpretations.  The  first term on the
right-hand  side   describes  the current   fluctuations  due  to  the
statistical   nature of the impurity scattering.    The next two terms
represent  fluctuations in    the diffusion current   associated  with
fluctuations   in the density   and  the  diffusion constant.    These
contributions are renormalized by the  quasiparticle interaction.  The
fourth  term reflect  fluctuations    in  the Lorentz force   due   to
fluctuations in the  Chern-Simons   magnetic  field.  This  term   was
previously discussed in  Ref.~\onlinecite{Ioffe}.  The  last two terms
describe  fluctuations in the response  to the electric  fields due to
fluctuations  in the   electric    fields    and the     conductivity,
respectively.

How are the current fluctuations  affected by the Chern-Simons fields?
To  answer  this question, it  is  instructive to derive the analogous
response equation  for classical electrons in a  magnetic field.  This
calculation is also done in appendix \ref{sec-diffusive} and one finds
\begin{eqnarray}
   {\bf\Delta j}={\bf\Delta J}-&&D^*{\bf\nabla_{\!r}}\Delta\rho-\Delta
     D^*{\bf\nabla_{\!r}}\rho   
    \nonumber\\    
     &&+\sigma{\bf\Delta    E}
     +\Delta\sigma{\bf E}-{e\tau_{\rm tr}\over m^*}({\bf B\times\Delta j}).
\label{response-B}
\end{eqnarray}
One observes that  the response equations  for composite  fermions and
electrons differ in two points:
\begin{itemize}
\item{The response equation for composite fermions includes  additional  
terms  involving  the  Chern-Simons  fields  and  their fluctuations.}
\item{The response equation for composite fermions lacks a Lorentz-force 
term due to the applied magnetic field.}
\end{itemize}
We will now  show that the  two response  equations are  equivalent to
leading order despite these seeming differences.

First,  we rewrite the  term in (\ref{response}) involving ${\bf\Delta
E}^{\rm CS}$ using Eq.\ (\ref{fluc-e}) and find
\begin{equation}
   \sigma_{\rm CF}{\bf\Delta E}^{\rm CS}=-{e\tau_{\rm tr}\over m^*}
     ({\bf B}_{1/2}{\bf\times\Delta j}),
\end{equation}
where it was used that  at half filling ${eB_{1/2}\tau_{\rm  tr}/m^*}=
-(2h/e^2)\sigma_{\rm CF}$.  Hence, this  term in the response equation
for   composite  fermions   reproduces  the   Lorentz-force  term  for
electrons.  This  is  analogous  to the  fact that   the  Chern-Simons
electric field leads to the  finite Hall resistivity in the derivation
of  the physical resistivity tensor from  the Boltzmann equation (cf.,
sec.\ \ref{sec-Boltzmann}).

We are now left with two additional terms in the response equation for
composite  fermions without  analog in   the case  of electrons  in  a
magnetic field.  We first discuss the  term involving the fluctuations
in  the conductivity. These  fluctuations arise due to fluctuations in
the density of the sample.   For a quadratic dispersion, the effective
mass is independent of the density so that
\begin{equation}
    \Delta\sigma_{\rm CF}={e\Delta\rho\,\tau_{\rm tr}\over m^*}.
\end{equation}
For arbitrary dispersions, there is an additional contribution arising
from fluctuations of the effective  mass with density.  Generally, the
terms  due  to   fluctuations  in  the  diffusion    constant  and the
conductivity are neglected  in the response  equation (\ref{response})
because  they are of order   ${\rm  max}\{T,eV\}/\mu$ relative to  the
leading contributions.  Here, it is  instructive to keep them  because
this allows one  to show that  the term involving ${\bf \Delta B}^{\rm
CS}$ is of the  same   order and hence   can  be neglected. In   fact,
replacing the Chern-Simons fields and $\Delta\sigma_{\rm CF}$ by their
explicit  expressions, one finds that  the  two terms actually  cancel
exactly for a quadratic dispersion,
\begin{equation}
   \Delta\sigma_{\rm CF}{\bf   E}^{\rm  CS}-{e\tau_{\rm tr}\over  m^*}
   ({\bf \Delta B}^{\rm CS}{\times j})=0.
\end{equation}
This shows that the term involving $\Delta{\bf B}^{\rm CS}$ is also of
order ${\rm max}\{T,eV\}/\mu$ relative to the leading terms and can be
neglected. Of  course, the  two terms  do  not cancel exactly   in the
general case of a non-quadratic dispersion.

The contribution of  ${\bf \Delta B}^{\rm  CS}$ to fluctuations in the
Hall voltage has been discussed in detail in Ref.\ \onlinecite{Ioffe}.
There it was argued that  these  Hall-voltage fluctuations could be  a
signature of the presence  of Chern-Simons fields  in the sample.  The
present approach shows that the dominant  contribution to Hall voltage
fluctuations does not come from ${\bf \Delta B}^{\rm CS}$, but instead
from    the   fluctuating Chern-Simons electric    field.   The latter
mechanism cannot be  used to verify the  presence of fluctuating gauge
fields  in the sample  because  it simply reflects   the presence of a
finite Hall resistivity, due  to which any current  fluctuation causes
fluctuations in the Hall voltage.

We briefly consider magnetic fields away from $\nu=1/2$. In this case,
the   semiclassical  approach   to   transport is  valid  as  long  as
Shubnikov-de   Haas oscillations   of    the  composite fermions   are
negligible.  Away  from $\nu=1/2$,  the  CF's experience  an  effective
magnetic field $B^*=B-B_{1/2}$    which leads to  the additional  term
$-(e\tau_{\rm tr}/m^*)({\bf  B^*\times  \Delta j})$  in  the  response
equation.  When  combined with  the  term due  to fluctuations  in the
Chern-Simons   electric field, one     obtains a   Lorentz-force  term
appropriate for the full externally applied magnetic field $B$.

As a result, we conclude that the density  and current fluctuations of
composite fermion in  the  diffusive regime do  not  differ to leading
order  from  the fluctuations of  classical  electrons in the external
magnetic field.   Neglecting  the  terms   involving $\Delta   D$  and
$\Delta\sigma$, we find for composite fermions near $\nu=1/2$
\begin{equation}
   {\bf\Delta j}={\bf\Delta
   J}-D^*{\bf\nabla_{\!r}}\Delta\rho+\sigma_{\rm
   CF}{\bf\Delta      E}   -{e\tau_{\rm tr}\over
   m^*}({\bf B\times \Delta j}).
\label{cf-result}
\end{equation}
Of course, the  analogy between the  response equations for  classical
electrons and  composite fermions concerns  the  form of  the response
equation.  The values of  the phenomenological constants such as $m^*$
and  $\tau_{\rm  tr}$  entering into   the  response equation  do  not
coincide.

\section{Low-frequency noise power}
\label{sec-zero-frequency}

In this section, the  general framework derived  above is  employed to
compute the  equilibrium and excess noise power  near $\nu=1/2$ in the
Corbino disc.  We will first consider the case of zero frequency.  The
question of finite frequency is discussed at the  end of this section.
The shot-noise power is computed for the two regimes $\ell_{\rm tr}\ll
L\ll L_{\rm cf-cf}\ll L_{\rm  cf-ph}$   and $\ell_{\rm tr}\ll   L_{\rm
cf-cf}\ll  L\ll L_{\rm   cf-ph}$.  For  samples   larger  than $L_{\rm
cf-ph}$, the shot-noise power  vanishes and there is only  equilibrium
noise.

For the  Corbino  disc we use a   coordinate system such that  the $x$
direction points in the radial direction  and the $y$ direction in the
angular direction  around  the  disc.  Then,   at  zero frequency, the
continuity equation implies that
\begin{eqnarray}
   \Delta  I_x&=&\int_{-L_y/2}^{L_y/2} dy\,\Delta j_x={1\over L_x}
     \int_\Omega dx\,dy\,\Delta j_x
     \\
   \Delta  I_y&=&\int_{-L_x/2}^{L_x/2} dx\,\Delta j_y={1\over L_y}
      \int_\Omega dx\,dy\,\Delta j_y.
\end{eqnarray}
Using the response equation (\ref{cf-result}) yields
\begin{eqnarray}
    \Delta I_x=&&{1\over L_x}\int_\Omega dx\,dy\,\Delta J_x
          +{2h\over e^2}\sigma_{\rm CF}{L_y\over L_x}\Delta I_y
        \nonumber\\
        &&-{1\over L_x}\int_{-L_y/2}^{L_y/2} dy\,\left[D^*\Delta\rho
        +\sigma_{\rm CF}\Delta\phi\right]^{x=L_x/2}_{x=-L_x/2}
\label{xresponse}
\end{eqnarray}
and
\begin{eqnarray}
    \Delta I_y=&&{1\over L_y}\int_\Omega dx\,dy\,\Delta J_y
          -{2h\over e^2}\sigma_{\rm CF}{L_x\over L_y}\Delta I_x
        \nonumber\\
        &&-{1\over L_y}\int_{-L_x/2}^{L_x/2} dx\,\left[D^*\Delta
        \rho+\sigma_{\rm CF}\Delta\phi\right]^{y=L_y/2}_{y=-L_y/2}.
\label{yresponse}
\end{eqnarray}
Here, we  introduced the  notation $[f(x)]_{x=a}^{x=b}=f(b)-f(a)$.  We
remark  that the  Hall  current around the  loop  can in  principle be
measured by  means of the  associated magnetic moment.  The quantities
in     the  square brackets       in   Eqs.\  (\ref{xresponse})    and
(\ref{yresponse})  are proportional  to the electrochemical  potential
differences across the    sample  in  the   $x$ and   $y$   direction,
respectively, if the distribution functions on the respective edges of
the  sample are the  equilibrium  distribution function.  This follows
from the standard result of Fermi-liquid theory\cite{Pines}
\begin{equation}
   \Delta\rho={\Delta\rho\over\Delta\mu}\Delta\mu={eN(0)\over1+F_0}
         \Delta\mu
\end{equation}
and the Einstein relation.

In the Corbino-disc geometry, the electrochemical potential difference
in the  $y$  direction (i.e., around   the  disc) must  vanish  due to
periodicity.  Hence, the last term in  Eq.\ (\ref{yresponse}) is zero.
Moreover,  since we  are  interested  in the  intrinsic  current noise
originating in the sample, we assume that the voltage source maintains
a  fixed electrochemical potential difference  across  the sample. For
the Corbino-disc  geometry,  this implies that also  the  last term in
Eq.\ (\ref{xresponse}) vanishes.

It is  now  simple  to  solve   the  equations (\ref{xresponse})   and
(\ref{yresponse}) for the current fluctuations,
\begin{eqnarray}
  \Delta I_x&&={1\over1+(2h\sigma_{\rm CF}/e^2)^2}
     \nonumber\\
      &&\times{1\over L_x}\int_\Omega dx\,dy
       \left\{\Delta J_x+(2h\sigma_{\rm CF}/e^2)^2\Delta J_y\right\}
       \\
  \Delta I_y&&={1\over1+(2h\sigma_{\rm CF}/e^2)^2}
      \nonumber\\
      &&\times{1\over L_y}\int_\Omega dx\,dy
       \left\{-(2h\sigma_{\rm CF}/e^2)^2\Delta J_x+\Delta J_y\right\}.
\end{eqnarray}
The associated zero-frequency noise powers are defined as
\begin{equation}
   S_\alpha=2\int dt\,\langle\Delta I_\alpha(t)\Delta I_\alpha
      (t^\prime)\rangle
\end{equation}
with $\alpha=x,y$. Using Eq.\ (\ref{deltaj}), one finds
\begin{eqnarray}
   S_x&=&4{\sigma_{\rm CF}L_y/L_x\over1+(2h\sigma_{\rm CF}/e^2)^2}
    \int_\Omega{d{\bf r}\over\Omega}\int d\epsilon\, n_\epsilon(1-n_\epsilon)
    \\
   S_y&=&4{\sigma_{\rm CF}L_x/L_y\over1+(2h\sigma_{\rm CF}/e^2)^2}
    \int_\Omega{d{\bf r}\over\Omega}\int d\epsilon\, n_\epsilon(1-n_\epsilon).
\label{shot-dist}
\end{eqnarray}
Note  that the noise power  $S_x$  of the longitudinal current between
edges differs from  the noise power $S_y$ of  the Hall  current around
the loop only by  geometrical factors.  The  prefactors are simply the
conductances of the Corbino disc.

If   the  composite-fermion   distribution  function   is  in  (local)
equilibrium,  we  can further  simplify the  expression  for the noise
power to
\begin{eqnarray}
   S_x&=&4{\sigma_{\rm CF}L_y/L_x\over1+(2h\sigma_{\rm CF}/e^2)^2}
    \int_\Omega{d{\bf r}\over\Omega}\, T_{\rm cf}({\bf r})
    \\
   S_y&=&4{\sigma_{\rm CF}L_x/L_y\over1+(2h\sigma_{\rm CF}/e^2)^2}
    \int_\Omega{d{\bf r}\over\Omega}\, T_{\rm cf}({\bf r})
\end{eqnarray}
in terms of  the local composite-fermion  temperature $T_{\rm cf}({\bf
r})$.  In  particular,  one immediately recovers from  this expression
the standard result for equilibrium noise.  Another simple application
are  samples   with $L\gg    L_{\rm   cf-ph}$.   In  this   case,  the
composite-fermion  temperature is   everywhere equal  to   the  phonon
temperature.  Hence, there is  only  equilibrium noise and shot  noise
vanishes.   Below,  we  will   derive the shot-noise  power   for  the
nontrivial regimes with $L\ll L_{\rm cf-ph}$.

Eq.\ (\ref{shot-dist})   expresses the zero-frequency  noise power  in
terms of the  isotropic   part of the average   distribution  function
$n_\epsilon$.   To linear   order   in the  applied  bias  and  in the
quasiparticle  interaction,   this   quantity  satisfies  a   modified
diffusion equation,
\begin{equation}
  D\nabla^2_{\!\bf r}n_{\epsilon-\delta(x,y)}+S^{\rm cf-cf}_{\bf p}
     \{n_{\bf p}\}=0.
\label{diffusion}
\end{equation}
where   $\delta(x,y)={\sum}_{\bf   p^\prime}  f_{\bf  pp^\prime}\delta
n_{\bf p^\prime}+e\phi$.   The quantity  $\delta$ depends only  on the
density distribution in  the sample and  is independent of $\epsilon$.
The derivation  of this    equation  from the  Boltzmann  equation  is
sketched   in  appendix    \ref{sec-average}.    The    solution    of
(\ref{diffusion}) simplifies for the  Corbino-disc geometry when using
the fact that $\sigma_{xy}/\sigma_{xx}\gg1$.  In this limit, the inner
and outer  edges  are, to a  good approximation,  equipotential lines,
even if the contacts  to the leads are local.   This implies  that the
distribution function becomes  essentially independent  of the angular
direction $y$.  With  this observation, we  can  state the appropriate
boundary conditions for  Eq.\ (\ref{diffusion}).  The battery provides
an electrochemical-potential difference $eV$ between the inner and the
outer edge,
\begin{equation}
   (\mu+e\phi)_{x=L_x/2}-(\mu+e\phi)_{x=-L_x/2}=eV.
\end{equation}
Hence, the chemical  potentials $\mu_i$ and $\mu_o$  at  the inner and
outer edges are
\begin{eqnarray}
  \mu_o&=&\mu-e\phi_o+{eV\over 2}\\
  \mu_i&=&\mu-e\phi_i-{eV\over 2},
\end{eqnarray}
with $\mu$ a constant. Since the distribution function in the leads is
in equilibrium, we have
\begin{equation}
  n_{\bf p}(x=\pm L_x/2)=f_{\mu\pm eV/2}(\tilde\epsilon_{\bf p}
    +e\phi(\pm L_x/2)),
\end{equation}
where $f_\mu(E)$ denotes the Fermi-Dirac distribution. From this, we 
obtain the boundary condition 
\begin{equation}
  n_{\epsilon-\delta(x=\pm L_x/2)}(x=\pm L_x/2)=f_{\mu\pm eV/2}(\epsilon)
\end{equation}
for the diffusion equation (\ref{diffusion}).

We first  specify to the limit of  weak CF-CF scattering, $L\ll L_{\rm
cf-cf}$, where     we  can neglect  the  collision    integral $S^{\rm
cf-cf}_{\bf p}$ in the diffusion equation (\ref{diffusion}). In this
limit, the distribution function becomes
\begin{eqnarray}
    n_\epsilon=&&\left[f_{\mu+eV/2}(\epsilon+\delta(x))-f_{\mu-eV/2}
     (\epsilon+\delta(x))\right]{x\over L}
     \nonumber\\
     &&+{1\over 2}
     \left[f_{\mu+eV/2}(\epsilon+\delta(x))+f_{\mu-eV/2}(\epsilon+\delta
     (x))\right].
\end{eqnarray}
Note that both the induced fluctuations in the physical electric field
and  the effects of  nontrivial Fermi-liquid  parameters are contained
entirely in the quantity $\delta(x)$.  Inserting this solution for the
average distribution function into Eq.\ (\ref{shot-dist}), one obtains
the final result
\begin{equation}
   S_\alpha={4 G_\alpha}\left\{{2\over3}T+{eV\over6}\coth{eV\over2T}
        \right\}.
\end{equation}
Here   we defined  the  conductances  
\begin{eqnarray}
   G_x&=&{\sigma_{\rm CF}L_y/L_x\over1+(2h\sigma_{\rm CF}/e^2)^2}
   \nonumber\\
   G_y&=&{\sigma_{\rm  CF}L_x/L_y\over
   1+(2h\sigma_{\rm   CF}/e^2)^2}
\end{eqnarray}
of the    Corbino disc.  Interestingly,  the   induced  electric field
fluctuations and nontrivial  Fermi-liquid parameters have no effect on
the noise power and  we obtain precisely  the same result for $S_x$ as
for usual diffusive contacts in the  absence of a  magnetic field.  In
the present problem,  there are additional  fluctuations $S_y$  in the
Hall current around  the loop due  to the finite Hall conductivity  of
the  sample.  The magnitudes of the   fluctuations in the longitudinal
and Hall currents differ only by geometric factors.

We  now turn to the  regime of  strong  CF-CF scattering, $L\gg L_{\rm
cf-cf}$.  In  this regime, CF-CF  scattering locally  equilibrates the
distribution  function so that $n_\epsilon$ can  be parameterized by a
local chemical potential    $\mu({\bf  r})$ and a    local temperature
$T_{\rm  cf}({\bf r})$,  which in general  does  not coincide with the
phonon temperature because of Joule heating.   An equation for $T_{\rm
cf}({\bf  r})$ can   be    derived   from the  diffusion      equation
(\ref{diffusion}).    Here  we follow    an  alternative, more  direct
approach.\cite{Steinbach}  The heat  current  carried by the composite
fermions due to a temperature gradient is
\begin{equation}
   {\bf j}^q=-\hat\sigma{\pi^2k_B^2\over6e^2}{\bf\nabla_{\!r}}
     T_{\rm cf}^2,
\label{heat-response}
\end{equation}
where $\hat\sigma$ denotes the   conductivity tensor.   In  principle,
there  is also  a  thermoelectric  contribution  to the heat  current,
which, however, can be neglected.  Joule heating acts  as a source for
the heat current so that
\begin{equation}
   {\bf\nabla_{\!r}j}^q=-{\bf j}^e {\bf\nabla_{\!r}}
       (\phi+{1\over e}\mu),
\label{heat-continuity}
\end{equation}
where ${\bf  j}^e$ denotes  the  charge current.  These equations  are
valid for arbitrary  Fermi-liquid  parameters.  For the  Corbino disc,
one readily finds that the Joule heating is
\begin{equation}
  -{\bf j}^e {\bf\nabla_{\!r}}(\phi+{1\over e}\mu)=\sigma_{xx}
     \left(V\over L_x\right)^2.
\end{equation}
Hence,    by        inserting     Eq.\   (\ref{heat-response})    into
(\ref{heat-continuity}),    we   obtain    an    equation    for   the
composite-fermion temperature,
\begin{equation}
   {\bf\nabla_{\!r}}^2T^2_{\rm cf}+{6e^2\over\pi^2k_B^2}\left(V\over 
      L_x\right)^2=0.
\label{el-temp}
\end{equation}
We assume that the   electrons are in  good  thermal contact with  the
phonon bath in the leads. Hence, we need to solve Eq.\ (\ref{el-temp})
subject  to  the   boundary condition  $T_{\rm   cf}(x=\pm  L_x/2)=T$.
Specifying for   simplicity  to $eV\gg   T$,  we  obtain  for  the  CF
temperature profile
\begin{equation}
   T_{\rm cf}(x)\simeq{\sqrt{3}\over2\pi}eV\sqrt{1-\left(2x\over
      L_x\right)^2}.
\end{equation}
Hence, the  noise  power in  the regime $L_{\rm  cf-cf}\ll L\ll L_{\rm
cf-ph}$ becomes
\begin{equation}
  S_\alpha={\sqrt{3}\over2}eG_\alpha V.
\end{equation}
The  result  for   $S_x$  is  again   the  same as   for  electrons in
zero-magnetic field,  when  written in  terms of  the average  current
flowing through the device.   The   fluctuations in the  Hall  current
differs  only  by   geometric factors from   the  fluctuations  of the
longitudinal current.

Finally, we address the question  of finite frequency.  Our  treatment
remains valid at  finite frequency as long  as the  time derivative of
the  density fluctuations can be  neglected in the diffusion equation.
For three-dimensional samples,  this is valid  for frequencies smaller
than   the  Maxwell  frequency  $4\pi\sigma_{xx}$.  In two-dimensional
systems, this frequency     is  smaller because screening    is   less
effective.  We estimate this frequency by considering
\begin{equation}
  {\bf k}\Delta{\bf j}={\bf k}{\hat\sigma}\Delta{\bf E}
          =-i2\pi\sigma_{xx}k(1/\epsilon)\Delta\rho,
\end{equation}
where we used Maxwell's  equation.  Hence, the zero-frequency   result
remains good for frequencies
\begin{equation}
   \omega\ll 2\pi\sigma_{xx}k/\epsilon\approx 10^8{\rm Hz}.
\end{equation}
The typical  scale for the  wavevector is set  by the system  size for
which we assumed  $10^{-4}$m for the numerical  estimate. We note that
the condition $\omega<1/\tau_{\rm tr}$ needed for  the validity of the
diffusive approximation is typically weaker than this requirement.

\section{Thermal correlators at finite wavevector and frequency}
\label{sec-thermal}

The Boltzmann-Langevin  formalism  can also    be applied to   compute
thermal   density-density and  current-current  correlators at  finite
frequency and wavevector  in the diffusive regime.   These correlators
describe  the  actual  density    and  current  fluctuations   in  the
sample. Once the correlators for the currents and densities are known,
the correlators  of the Chern-Simons electric  and magnetic fields and
for  the  physical    electric   field can   be    obtained from Eqs.\
(\ref{CS-magnetic}), (\ref{electric}), and (\ref{CS-electric}).

Writing   the response  equation  (\ref{cf-result})  in  frequency and
momentum space, we have
\begin{equation}
   {\bf\Delta j}={\bf\Delta J}-iD^*{\bf k}\Delta\rho
     +\sigma_{\rm CF}{\bf\Delta E}
      -{e\tau_{\rm tr}\over m^*}({\bf B\times \Delta j}).
\end{equation}
The density and  electric-field  fluctuations can be  eliminated  from
this equation using the continuity equation
\begin{equation}
   \omega\Delta\rho={\bf k\Delta j}
\end{equation}
and Maxwell's equation
\begin{equation}
   \epsilon{\bf\Delta E}=-i2\pi{\bf\hat k}\Delta\rho,
\end{equation}
where $\hat{\bf k}$ is the unit vector in  the direction of ${\bf k}$.
Decomposing the current into its components parallel and perpendicular
to  the  wavevector $\Delta   j_\parallel={\bf   \hat k\Delta  j}$ and
$\Delta j_\perp={\bf \hat k(\hat z\times\Delta j)}$, one finds
\begin{equation}
    \Delta j_\parallel=\Delta J_\parallel-i{D^*k^2\over\omega}
    \Delta j_\parallel-i{2\pi\sigma_{\rm CF}|k|\over\epsilon\omega}
    \Delta j_\parallel+{\omega_c\tau_{\rm tr}}\Delta j_\perp
\end{equation}
and
\begin{equation}
  \Delta j_\perp=\Delta J_\perp-{\omega_c\tau_{\rm tr}}\Delta j_\parallel.
\end{equation}
These equations are readily solved and one obtains
\begin{equation}
  \Delta j_\parallel={\Delta J_\parallel+{\omega_c\tau_{\rm tr}}\Delta J_\perp
      \over 1+({\omega_c\tau_{\rm tr}})^2+(i/\omega)[D^*k^2+2\pi
      \sigma_{\rm CF}|k|/\epsilon]}
\end{equation}
and
\begin{eqnarray}
  \Delta j_\perp&=&\Delta J_\perp
    \nonumber\\
    &&-{{\omega_c\tau_{\rm tr}}(\Delta J_\parallel+
     {\omega_c\tau_{\rm tr}}\Delta J_\perp)\over 1+({\omega_c\tau_{\rm tr}})^2
     +(i/\omega)[D^*k^2+2\pi\sigma_{\rm CF}|k|/\epsilon]}.
\end{eqnarray}
From the definition of the source current $\Delta{\bf J}$, one obtains
in thermal equilibrium 
\begin{eqnarray}
  \langle\Delta J_\parallel\Delta J_\parallel\rangle_{\omega,\bf k}
   &=&\langle\Delta J_\perp\Delta J_\perp\rangle_{\omega,\bf k}
    =2T\sigma_{\rm CF}
    \nonumber\\
   \langle\Delta J_\parallel\Delta J_\perp\rangle_{\omega,\bf k}&=&0.
\end{eqnarray}
Here we use  the  notation $\langle fg\rangle_{\omega,\bf k}  =\langle
f({\omega,\bf  k})g({-\omega,-\bf   k})\rangle$.    This yields    the
current-current correlators
\begin{equation}
  \langle\Delta j_\parallel\Delta j_\parallel\rangle_{\omega,\bf k}
    ={2T\sigma_{xx}\omega^2\over\omega^2+[D^*_{xx}k^2+2\pi\sigma_{xx}k
   /\epsilon]^2},
\end{equation}
and
\begin{eqnarray}
  \langle\Delta j_\perp\Delta j_\perp&&\rangle_{\omega,\bf k}
    =2T\sigma_{xx}
   \nonumber\\
   &&\times\left\{1+{[D^*_{xy}k^2+2\pi\sigma_{xy}k]^2
   \over\omega^2+[D^*_{xx}k^2+2\pi\sigma_{xx}k/\epsilon]^2}\right\}.
\end{eqnarray}
Here, $\sigma_{\alpha\beta}$  denote the  components  of the  physical
conductivity  tensor  and    we  defined   a  diffusion   tensor    by
$\sigma_{\alpha\beta}=e^2N(0)D_{\alpha\beta}$.     Note   that   these
correlators also corroborate   the  estimate for the  validity  of the
zero-frequency approximation at   the end  of the  previous   section.
Using the  continuity  equation,  one  finds for   the density-density
correlator
\begin{equation}
  \langle\Delta\rho\Delta\rho\rangle_{\omega,\bf k}
    ={2T\sigma_{xx}k^2\over\omega^2+[D^*_{xx}k^2+2\pi\sigma_{xx}k
    /\epsilon]^2}.
\end{equation}
All of these correlators for composite fermions are identical to those
for semiclassical electrons in  the external magnetic field.  From the
Boltzmann-Langevin  approach,  this result   follows directly  for the
thermal correlators.    The analogous result   for  the  retarded  and
advanced correlators follows  from the fluctuation-dissipation theorem
combined with the Kramers-Kronig relations.

\section{Conclusions and summary}
\label{sec-conclusion}

In this paper,  we have  studied  non-equilibrium density  and current
fluctuations   in  the   half-filled    Landau  level,  employing  the
composite-fermion picture and focusing on   the diffusive regime.   By
deriving   a  Boltzmann-Langevin   equation for  composite   fermions,
including both the coupling  to the Chern-Simons fields and  arbitrary
quasiparticle interactions, we found  that, {\it to leading order, the
density and current fluctuations for  diffusive samples near $\nu=1/2$
are equivalent  to those of  semiclassical electrons in the externally
applied  magnetic  field.}     Current  fluctuations  associated  with
fluctuations  in  the Chern-Simons  magnetic   field are suppressed by
${\rm    max}\{eV,T\}/\mu$  relative   to  the leading  contributions.
Fluctuations in the  Chern-Simons  electric  field are  important  and
reproduce the   term arising  due    to the large  Lorentz force   for
semiclassical electrons.  One consequence of this is that fluctuations
in the Hall voltage are dominated by  fluctuations in the Chern-Simons
electric field, reflecting the large  Hall conductivity of the sample,
rather than fluctuations in the Chern-Simons  magnetic field.  This is
in contrast to previous suggestions.\cite{Ioffe}

The general results for density  and current fluctuations can be  used
to compute the shot-noise power in  the half-filled Landau level.  For
the Corbino-disc   geometry,  we found that   the shot-noise  power at
$\nu=1/2$, when expressed in    terms of the average current   flowing
through the device,  equals that for  metallic systems in low magnetic
fields.  This implies that diffusive shot noise near $\nu=1/2$ remains
unaffected by quasiparticle  interactions and by  the coupling to  the
Chern-Simons fields.    We note in  passing that  the insensitivity of
shot noise to Fermi-liquid corrections also holds for regular metallic
contacts.  Shot noise in a two-terminal conductor near $\nu=1/2$ would
also   be  interesting       for   comparison    with   results    for
fractional-quantum-Hall    states,\cite{Kane}  but  the  corresponding
calculations are complicated by the  fact that, in this geometry, most
of  the resistance is associated with  the contacts between sample and
leads.  Finally,  we find  that shot noise  in  the half-filled Landau
level depends sensitively on the ratio of the sample size to the CF-CF
and the CF-phonon scattering lengths.   Our results suggest that  shot
noise can be used to measure these important length scales.

\acknowledgments
I would like  to acknowledge support by a  scholarship of  the Minerva
Foundation, Munich,  a Minerva project  grant, and grant no.\ 95-250/1
of the  U.S.-Israel  Binational Science  Foundation.  I  enjoyed  very
useful and  informative discussions   with  Rafi de-Picciotto,   Misha
Reznikov, and Ady Stern.

\appendix

\section{Noise in diffusive systems in presence of magnetic field}
\label{sec-diffusive}

In this  appendix we provide details of   the calculations sketched in
Sec.\ \ref{sec-BL}.   In  particular, we  will show how  to derive the
continuity and response equations from the Boltzmann-Langevin equation
including  the terms   arising   from fluctuations   in  the diffusion
constant which are usually neglected but turned out to be conceptually
important  in   the present context.   The  present   calculation also
includes the effects of a non-vanishing quasiparticle interaction.  We
will restrict our calculation to  strictly two-dimensional samples and
consider a system of classical electrons in the diffusive regime.  The
derivation of the response   equation (\ref{response}) for   composite
fermions at $\nu=1/2$ is completely analogous.

The Boltzmann-Langevin equation is 
\begin{eqnarray}
  {\partial\over\partial t}\Delta n_{\bf p}+{\bf v_p\nabla_{\!r}}
     &&\Delta\tilde n_{\bf p}+e{\bf E\nabla_{\!p}}\Delta n_{\bf p}
     +e({\bf v_p\times B}){\bf\nabla_{\!p}}\Delta\tilde n_{\bf p}
     \nonumber\\
     +e\Delta{\bf E\nabla_p}n_{\bf p}&&
     -S^\prime_{\bf p}\{\Delta n_{\bf p}\}=\Delta J_{\bf p}.
\end{eqnarray}
We  start by  introducing   the decompositions (\ref{decom-dist})  and
(\ref{decom-source}) into the Boltzmann-Langevin equation.  Collecting
the resulting terms  in   the Boltzmann-Langevin equation which    are
isotropic in $\bf p$, we find
\begin{eqnarray}
  {\partial\over\partial t}\Delta n_\epsilon+{\bf v_p\nabla_{\!r}}
    ({\bf v_p\Delta\tilde f}_\epsilon)
   +e{\bf E}{\bf\nabla_{\!p}}({\bf v_p\Delta f}_\epsilon)
     &&
     \nonumber\\
    +e({\bf v_p\times B}){\bf\nabla_{\!p}}\Delta\tilde n_\epsilon
     +e\Delta{\bf E\nabla_p}({\bf v_p f}_\epsilon)
      &&=0.
\label{isotropic}
\end{eqnarray}
Here, we used that $\Delta J_\epsilon$ has  zero average and variance.
To derive an  equation in terms  of the  charge and  current densities
(\ref{density}) and (\ref{current}),  we   multiply this equation   by
$e/\Omega$ and sum  over all  ${\bf p}$.  We   will now show  that the
equation     will   then     reduce    to   the    continuity equation
(\ref{continuity}).  One immediately finds that the first two terms in
(\ref{isotropic}) give  the time  derivative of  the  density  and the
divergence  of the current  density.  The electric-field term vanishes
because
\begin{eqnarray}
      {e^2\over\Omega}\sum_{\bf p}&&{\bf E}{\bf\nabla_{\!p}}({\bf v_p
      \Delta f}_\epsilon)
      \nonumber\\
      &&={e^2\over\Omega}\sum_{\bf p}{\bf E}\left\{{{\bf\Delta f}_\epsilon
      \over m^*}+{\bf v_p}\left({\bf v_p}{\partial{\bf \Delta f}_\epsilon
      \over\partial\epsilon}\right)\right\}
      \nonumber\\
      &&={e^2\over m^*}N(0)\int d\epsilon\,{\bf E}\left\{{\bf\Delta 
      f}_\epsilon+\epsilon{\partial{\bf\Delta f}_\epsilon\over\partial
      \epsilon}\right\}
      \nonumber\\
      &&=0
\end{eqnarray}
The term involving the fluctuations of the electric field vanishes for
the same reason.   Finally, the magnetic-field term obviously vanishes
because of the vector structure. This completes  the derivation of the
continuity equation from the Boltzmann-Langevin equation.

Next, we collect the  terms  in the Boltzmann-Langevin  equation which
are anisotropic in the momentum $\bf p$,
\begin{eqnarray}
  {\bf v_p\nabla_{\!r}}\Delta\tilde n_\epsilon+e{\bf E}{\bf\nabla_{\!p}}
   \Delta n_\epsilon+e({\bf v_p\times B}){\bf\nabla_{\!p}}({\bf v_p 
   \Delta\tilde f}_\epsilon)&&
   \nonumber\\
   +e\Delta{\bf E\nabla_p}n_\epsilon+{1\over\tau_{\rm tr}}({\bf v_p 
   \Delta\tilde f}_\epsilon)=\Delta && J_{\bf p}.
\label{anisotropic}
\end{eqnarray}
Here,  we neglected  the time   derivative  relative to  the collision
integral.   To  derive  the response equation   (\ref{response-B}), we
multiply by $(e\tau_{\rm tr}/\Omega){\bf  v_p}$ and sum over all ${\bf
p}$.      We will again   consider   each term   separately.  The term
originating from the collision integral gives the current fluctuations
$\bf\Delta  j$.   The remaining terms  on  the  left-hand-side of Eq.\
(\ref{anisotropic}) require some more work.

The spatial gradient term  yields the two terms  in (\ref{response-B})
which  are associated   with fluctuations in  the diffusion  currents.
Performing the integral over the directions of $\bf p$, we have
\begin{equation}
    {e\tau_{\rm tr}\over\Omega}\sum_{\bf p}{\bf v_p}({\bf v_p\nabla_{\!r}}
       \Delta\tilde n_\epsilon)={e\tau_{\rm tr}\over m^*}N(0){\bf\nabla_{\!r}}
       \int d\epsilon\,\epsilon\Delta\tilde n_\epsilon.
\end{equation}
From the definition
\begin{equation}
  \Delta\tilde n_{\bf p}=\Delta n_{\bf p}-{\partial n^0_{\bf p}\over
    \partial\epsilon_{\bf p}}\sum_{\bf p^\prime}f_{\bf pp^\prime}
     \Delta n_{\bf p^\prime},
\end{equation}
we find
\begin{equation}
   \Delta\tilde n_\epsilon=\Delta n_\epsilon-{\partial n^0_\epsilon\over
    \partial\epsilon} f_0{\Omega\over e}\Delta\rho.
\end{equation}
Here, we introduced   the angular average  $f_0$  of the quasiparticle
interaction  function $f_{\bf   pp^\prime}$.  The fluctuations  of the
distribution function occur in a narrow energy window around the Fermi
energy. Hence, we can approximate
\begin{equation}
   \Delta n_\epsilon=-A\,{\partial n^0_\epsilon\over\partial\epsilon},
\end{equation}
with
\begin{equation}
   A={1\over eN(0)}\Delta\rho.
\end{equation}
Hence,
\begin{equation}
   \Delta\tilde n_\epsilon=-{1\over eN(0)}{\partial n^0_\epsilon\over
     \partial\epsilon}(1+F_0)\Delta\rho.
\end{equation}
with the Landau parameter  $F_0=\Omega N(0)f_0$. Using  this relation,
we obtain for the spatial gradient term
\begin{eqnarray}
   {e\tau_{\rm tr}\over\Omega}\sum_{\bf p}&&{\bf v_p}({\bf v_p\nabla_{\!r}}
       \Delta\tilde n_\epsilon)
   ={\tau_{\rm tr}\over em^*N(0)}(1+F_0){\bf\nabla_{\!r}}\{\rho\,\Delta\rho\}
   \nonumber\\
   &&=D(1+F_0){\bf\nabla_{\!r}}\Delta\rho
     +\Delta D(1+F_0){\bf\nabla_{\!r}}\rho.
\end{eqnarray}
with the diffusion-constant fluctuations  given by $\Delta D=\tau_{\rm
tr}\Delta\rho/em^*N(0)$.

The electric-field term becomes
\begin{eqnarray}
     {e^2\tau_{\rm tr}\over\Omega}\sum_{\bf p}{\bf v_p}({\bf E}
     {\bf\nabla_{\!p}})\Delta n_\epsilon
     &&={e^2\tau_{\rm tr}\over m^*}{\bf E}N(0)\int d\epsilon\,
     \epsilon{\partial\Delta n_\epsilon\over\partial\epsilon}
     \nonumber\\
     &&=-{e\tau_{\rm tr}\over m^*}\Delta\rho {\bf E}
     \nonumber\\
     &&=-\Delta\,\sigma{\bf E}.
\end{eqnarray}
with  $\Delta\sigma=e^2N(0)\Delta D$.   By  analogous steps, one finds
that the term involving fluctuations in the electric field becomes
\begin{equation}
 {e^2\tau_{\rm tr}\over\Omega}\sum_{\bf p}{\bf v_p}({\bf\Delta E}
    {\bf\nabla_{\!p}})n_\epsilon=-\sigma{\bf\Delta E}.
\end{equation}
This completes the derivation of the terms involving the electric field 
in (\ref{response-B}).

The magnetic-field term is evaluated as 
\begin{eqnarray}
   {e\tau_{\rm tr}\over\Omega}e\sum_{\bf p}&&{\bf v_p}\left[({\bf v_p\times B})
    {\bf\nabla_{\bf p}}({\bf v_p\Delta\tilde f}_\epsilon)\right]
    \nonumber\\
     &&={e^2\tau_{\rm tr}\over (m^*)^2\Omega}N(0)\int d\epsilon\,\epsilon
      ({\bf B\times \Delta\tilde f}_\epsilon)
     \nonumber\\
     &&={e\tau_{\rm tr}\over m^*}({\bf B\times\Delta j}),
\end{eqnarray}
where I used 
\begin{eqnarray}
  \Delta {\bf j}&=&{e\over\Omega}\sum_{\bf p}{\bf v_p}
     ({\bf v_p\Delta\tilde f}_\epsilon)
     \nonumber\\
     &=&{e\over m^*\Omega}N(0)\int d\epsilon\,\epsilon\Delta{\bf\tilde 
      f}_\epsilon.
\end{eqnarray}
This      completes  the  derivation        of the response   equation
(\ref{response-B}) of   classical  diffusive electrons in   a magnetic
field.

\section{Average distribution function}
\label{sec-average}

In this  appendix,  we  provide  details  of   the derivation of   the
diffusion  equation  (\ref{diffusion}) for the  isotropic  part of the
average  distribution    function   from   the    Boltzmann   equation
(\ref{Boltzmann}). We decompose the average distribution function into
its isotropic and anisotropic parts (in momentum $\bf p$),
\begin{eqnarray}
    n_{\bf p}&=&n_\epsilon+{\bf v_p f}_\epsilon
    \label{decom-av-dist}
     \\
     J_{\bf p}&=&J_\epsilon+{\bf v_p J}_\epsilon.
     \label{decom-av-source}
\end{eqnarray}
Inserting   this  decomposition     into  the  Boltzmann      equation
(\ref{Boltzmann}) and collecting    terms  which  are  isotropic   and
anisotropic in the momentum $\bf p$, we obtain the two equations
\begin{equation}
   ({\bf v_p\nabla_{\!r}}){\bf v_p\tilde f_\epsilon}-S^{\rm cf-cf}_{\bf p}
     (n_{\bf p})=0
\end{equation}
and
\begin{equation}
  ({\bf v_p\nabla_{\!r}})\tilde n_\epsilon+e({\bf E+E}^{\rm CS}){\bf v_p}
   {\partial n^0_\epsilon\over\partial\epsilon}+{1\over\tau_{\rm tr}}
   {\bf v_p\tilde f_\epsilon}=0.
\end{equation}
Inserting   the  second  into  the  first   equation,  using  that the
divergence of the Chern-Simons electric field vanishes, and expressing
the physical  electric  field in terms   of the scalar  potential, one
finds
\begin{equation}
   D\nabla^2_{\!\bf r}\left\{\tilde n_\epsilon-e\phi{\partial 
   n^0_\epsilon\over\partial\epsilon}\right\}+S^{\rm cf-cf}_{\bf p}
     (n_{\bf p})=0.
\end{equation}
Using the relation\cite{Pines}
\begin{equation}
   \tilde n_\epsilon=n_\epsilon-{\partial 
     n^0_\epsilon\over\partial\epsilon}\sum_{\bf p^\prime}
     f_{\bf pp^\prime}\delta n_{\bf p^\prime},
\end{equation}
we obtain  Eq.\ (\ref{diffusion}) within  the accuracy of Fermi-liquid
theory.


\begin{references}

\bibitem{Jain} J.K.\ Jain, Phys.\  Rev.\ Lett.\ {\bf 63}, 199  (1989);
Phys.\  Rev. B {\bf 40}, 8079  (1989); Phys.\ Rev.\  B  {\bf 41}, 7653
(1990); Adv.\ Phys.\ {\bf 41}, 105 (1992).

\bibitem{Halperin} B.\ Halperin, P.A.\ Lee, and N.\ Read, Phys.\ Rev.\ 
  B {\bf 47}, 7312 (1993).

\bibitem{Halperin-review} For a review, see B.\ Halperin in
{\it New  perpectives in quantum Hall effects},''  ed.\ by S. Das Sarma
and A.\ Pinczuk (Wiley \& Sons, 1997).

\bibitem{Lopez} A.\ Lopez and E.\  Fradkin,  Phys.\ Rev.\ B {\bf  44},
   5246 (1991); Phys.\ Rev.\ B {\bf 47}, 7080 (1993). 

\bibitem{Simon} S.H.\ Simon and B.I.\ Halperin, Phys.\ Rev.\ B {\bf 48},
17368 (1993).

\bibitem{Kim} Y.B.\ Kim, A.\ Furusaki, X.-G.\ Wen, and P.A.\ Lee, 
Phys.\ Rev.\ B {\bf 50}, 17917 (1994). 

\bibitem{Stern} A.\ Stern and B.I.\ Halperin, Phys. Rev.\ B {\bf 52},
5890 (1995).

\bibitem{Review} For a recent review of shot noise, see  M.J.M.\ de 
   Jong and C.W.J.\ Beenakker, Report No.\ cond-mat/9611140

\bibitem{Reznikov} M.\  Reznikov, M.\ Heiblum,  H.\ Shtrikman, and D.\
  Mahalu, Phys.\ Rev.\  Lett.\ {\bf 75},  3340 (1995); A.\  Kumar, L.\
  Saminadayar,  D.C.\ Glattli, Y.\ Yin,  and B.\ Etienne, Phys.\ Rev.\
  Lett.\ {\bf 76}, 2778 (1996).
  
\bibitem{Steinbach} A.H.\ Steinbach, J.M.\ Martinis, and M.H.\ Devoret,
  Phys.\ Rev.\ Lett.\ {\bf 76}, 3806 (1996).

\bibitem{Lesovik}  G.B.\ Lesovik, JETP  Lett.   {\bf 49}, 592  (1989).
  
\bibitem{Beenakker} C.W.J.\ Beenakker  and M. B\"uttiker, Phys.\  
  Rev.\ B {\bf 46}, 1889(1992).
  
\bibitem{Nagaev1} K.E.\ Nagaev, Phys.\ Lett.\ A {\bf 169}, 103 (1992).

\bibitem{Altshuler} B.L.\ Altshuler, L.\ Levitov, and A.Yu.\ Yakovets,
  JETP Lett. {\bf 59}, 857 (1994).

\bibitem{Nagaev2} K.E.\ Nagaev, Phys. Rev. B {\bf 52}, 4740 (1995).

\bibitem{Kozub} V.I.\ Kozub and A.M.\ Rudin, Phys. Rev. B {\bf 52}, 
  7853 (1995).

\bibitem{Kogan} Sh.M.\ Kogan and A.Ya.\ Shul'man, Sov.\ Phys.\ JETP
  {\bf 29}, 467 (1969).

\bibitem{Kang} W.\ Kang, S.\ He, H.L.\ Stormer, L.N.\ Pfeiffer, 
K.W.\  Baldwin, and  K.W.\ West,  Phys.\  Rev.\ Lett.\  {\bf 75}, 4106
(1995); it can  be deduced from  this experiment that $L_{\rm  cf-ph}$
becomes orders of magnitude larger than $\ell_{\rm tr}$ at $T\ll 1$K.

\bibitem{Pines} D.\ Pines and P.\ Nozi\`eres, {\it The theory of 
quantum liquids},  (Addison-Wesley, 1989); we   follow the notation of
this book.

\bibitem{relax} Strictly speaking, the relaxation-time approximation
is not   justified  for the  magnetic  scatterers  which dominate  for
composite  fermions [see e.g., A.D.\ Mirlin  and  P.\ W\"olfle, Phys.\
Rev.\ Lett.\  {\bf 78}, 3717 (1997)].   However,  one can  show that a
more realistic collision integral would not  change our results, since
we work in the diffusive regime.

\bibitem{Ioffe} L.B.\ Ioffe, G.B.\ Lesovik, and A.J.\ Millis, 
  Phys.\ Rev.\ Lett.\ {\bf 77}, 1584 (1996).

\bibitem{Kane} C.L.\ Kane and M.P.A.\ Fisher, Phys.\ Rev.\ Lett.\ 
  {\bf 72}, 724 (1994); P.\ Fendley, A.W.W.\ Ludwig, and H. Saleur, 
  Phys.\ Rev.\ Lett.\ {\bf 75}, 2196 (1995).

\end{references}
\end{document}